# DEVELOPMENT OF LIQUID XENON DETECTORS FOR MEDICAL IMAGING

V.CHEPEL, M.I.LOPES, V.SOLOVOV, R.FERREIRA MARQUES, A.J.P.L.POLICARPO

*LIP-Coimbra and Department of Physics of the University of Coimbra, 3004-516 Coimbra, Portugal*

In the present paper, we report on our developments of liquid xenon detectors for medical imaging, positron emission tomography and single photon imaging, in particular. The results of the studies of several photon detectors (photomultiplier tubes and large area avalanche photodiode) suitable for detection of xenon scintillation are also briefly described.

## 1   Introduction

Liquid xenon is well known to be an excellent detecting medium for γ-rays due to its high atomic number and density, fast scintillation and high scintillation light yield comparable to that of NaI(Tl) – the best known scintillator, and the fact that the ionisation electrons can be easily extracted from the particle track and, thus, give rise to a charge signal. The potential of liquid xenon detectors for medical imaging with radioisotopes has been recognized in early 70-s when several pioneering works have been published. The BNL group tested a multiwire proportional chamber filled with liquid xenon to detect 140 keV γ-rays reading each wire with an individual low-noise preamplifier, very expensive at that time [1]. The multiplication gain was only about 10. To overcome this problem, the wires were put into the gas phase above the liquid and, thus, the electrons were transported by the electric field from the liquid to gas [2]. In this two-phase version, higher gain could be reached but still not enough for using conventional electronics. The gain was limited by instabilities due to photon feedback. Alternatively, Egorov et al [3] developed a two-phase scintillation gamma camera, in which the signal was amplified by accelerating the electrons in a uniform electric field in the gas so that the secondary scintillation light was produced. The photon amplification factor of about several hundreds with respect to primary scintillation was achieved. The light was detected with an array of photomultipliers connected to an Anger scheme for determining the position in two dimensions.

The advantages of using liquid xenon for positron emission tomography (PET) were pointed out in the work of Lavoie [4]. He suggested to use the two virtues of liquid xenon - scintillation and charge – for fast timing and precise positioning, respectively. However, the problem of detection of small charge signals remained.

Significant progress in electronics in the past decades makes nowadays feasible to think about many thousands readout channels with very low noise front-end pre-amplification stage for quite low price per channel. It turns out that it is not really necessary to multiply electrons in the liquid (or in the gas phase) to measure the ionisation signals from γ-rays with the energy above ~100 keV. The noise of an "average price" commercially available charge sensitive preamplifier (not the best one) can be as low as 300 electrons, r.m.s., that allows the signals with about 5,000 electrons, corresponding to the above γ-ray energy, to be reliably measured. The possibility of intrinsic amplification of the ionisation signal in the liquid, even if the gain is not very high, is still attractive as it can allow further simplification of the readout electronics. Therefore, new attempts were undertaken with recently developed microstructures [5], but little practical advancement has been made so far.

PET detectors, which rely only on liquid xenon scintillation, were also recently suggested [6,7].

The new situation in what concerns low noise electronics opened up new prospects for liquid xenon ionisation detectors, for medical imaging, in particular. In the present paper we describe our work on the development of a detector module for PET, which makes use both of scintillation and ionisation signals in the liquid, and an ionisation chamber with 2D position sensitivity for single photon imaging. We also describe shortly our results on low temperature tests of VUV sensitive photomultiplier tubes and photodiodes in view of their use for the detection of scintillation of liquid xenon.

## 2 Study of VUV photon detectors

Liquid xenon scintillates in the vacuum ultraviolet wavelength region with $\lambda_{max} \approx 175$ nm. This rises two difficulties, one being the light collection due to poor reflectivity of most materials in VUV and the other the photon detection. The photons are usually detected with photomultiplier tubes (PMT) with a VUV transparent window. In order to reduce photon losses, the tube should be immersed into the liquid, i.e. to operate at the liquid xenon temperature. However, the resistivity of the photocathode materials increases with decreasing temperature and this can result in the tube failure. This effect is most significant for bialkaline photocathodes that are of much interest due to their high quantum efficiency and low dark noise. Therefore, we studied the behavior of several PMT models with bialkaline, multialkaline and CsSb photocathodes at temperatures down to -170ºC paying special attention to the bialkaline PMTs. The tests were carried out both in DC and pulse mode with different wavelengths. The illumination intensity varied between ~$10^3$ to ~$10^5$ photons per pulse or, in DC mode, from 14.6 pW/cm$^2$ to 280 pW/cm$^2$ that corresponds to ~$10^2$ and ~$2 \times 10^3$ photons/μs for the whole photocathode area, respectively. In the DC mode, the photocurrent was measured directly at the photocathode. The count rate and the illuminated area were also varied [8,9].

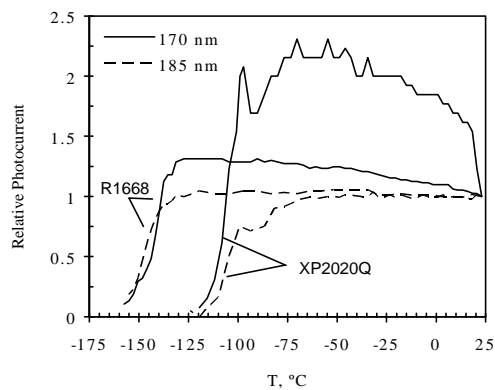

**Figure 1.** Relative photocurrent as a function of temperature for two models of PMTs with bialkaline photocathodes in VUV (Hamamatsu R1668, 1-1/8" diameter, and 2" Philips XP2020Q).

All the tested photomultipliers show dependence of their response (photocurrent or anode amplitude for pulse mode) on the temperature. Fig.1 shows this dependence for two PMT models and two wavelengths. One can distinguish two regions: the one where the current varies more or less smoothly and the other where it abruptly drops practically to zero. The smooth variation depends on the wavelength and is attributed to the shift of the photocathode spectral sensitivity to shorter $\lambda$. The unusually large effect at $\lambda=170$ nm for XP2020Q does not seem to fit to the above explanation. A somewhat similar behaviour was found also for EMI9750Q. This effect may be responsible for the discrepancies in the published data on the energy expended per scintillation photon in liquid xenon.

The abrupt drop can be characterised by a critical temperature, $T_c$, which depends on the illumination intensity and count rate (the higher the illumination/count rate, the higher the critical temperature) and is practically independent on $\lambda$. If only a limited region of the photocathode is illuminated, $T_c$ almost does not change but the drop of photocurrent becomes much faster. The photocathode diameter also affects $T_c$.

The sharp drop at $T_c$ is very well described in terms of photocathode resistivity with a simple model developed in [9], which allows to predict $T_c$, in some cases. The increase of the resistivity with decrease of the temperature leads to charging up the photo-emissive layer and thus prevents the photoelectron emission. Therefore, the deposition of a metal underlay on the PMT window seems to be a solution for the problem.

Recently developed and made commercially available Large Area Avalanche Photodiodes can be a very attractive alternative to photomultiplier tubes as they have high quantum efficiency, are compact, do not need voltage divider (but need low noise preamplifier). For low background experiments, the additional benefits are small mass and low intrinsic radioactivity. We have tested a 5 mm diameter photodiode immersed into liquid xenon. It was found that its quantum efficiency is ~100% at 175 nm, it is as fast as PMT (time resolution of 0.9 ns (*fwhm*) was measured for 511 keV γ-rays in coincidence with $BaF_2$), the dark current decreases by a factor of ~$10^5$ with respect to room temperature and corresponds to a few electrons for 1 μs shaping time at –110ºC. The energy resolution of 10% (*fwhm*) was obtained with 5.5 MeV α-particles at 15 mm from the photodiode. We refer to [10,11] for further details.

## 3  PETYA – a new detector for PET

### 3.1  Requirements to a PET detector

In PET, a radioactive tracer labeled with a positron emitting isotope is introduced into the object of study. After slowing down, the positron annihilates with an electron thus resulting in two back-to-back γ-photons of 511 keV each, almost collinear. These two photons, if detected and the interaction position within the detector measured for each of them, define a straight line (chord) that passes through the source. Accumulating a large number of such lines allows the image of an extended source to be reconstructed. Therefore, the detector must have two-dimensional position sensitivity, at least, with the resolution of the order of one or a few mm, depending on the application. However, 3D position sensitivity is very much preferable because it would allow, first, to use also the oblique chords in the reconstruction (and, thus, increase the system sensitivity) and, second, to account for the parallax, which appears when the source is displaced from the ring center.

If one or both γ-rays were scattered in the object, the obtained chord would not pass through the emission point and, therefore, such event should be discarded. This is possible if the detector can also measure the energy of the γ-rays. The existing detectors have energy resolution of ~20% for 511 keV.

Time resolution is also important as a narrow time coincidence window allows the rate of random coincidences to be reduced. The count rate capability should be ~$10^4$ to ~$10^5$ $s^{-1}cm^{-2}$ to provide the necessary statistical image quality at reasonable exposition time. Finally, the detection efficiency as high as possible is required as it is one of the factors that determine the dose given to patient.

### 3.2  Design

In order to satisfy high counting rate requirement to a PET detector, the electron collection time should be as short as possible. This leads to a multi-sectional concept of the chamber [12]. Fig.2 depicts schematically the design of the chamber that we tested. It is composed by an array of cells of 1 cm width separated by thin cathode plates with a plane of collecting wires in the middle. The cells are 60 mm long and 50 mm high (along the direction of incoming γ-rays). For this geometry, the probability for a 511 keV γ-ray to interact with liquid xenon is about 73% and the maximum electron collection time is 2.2 µs.

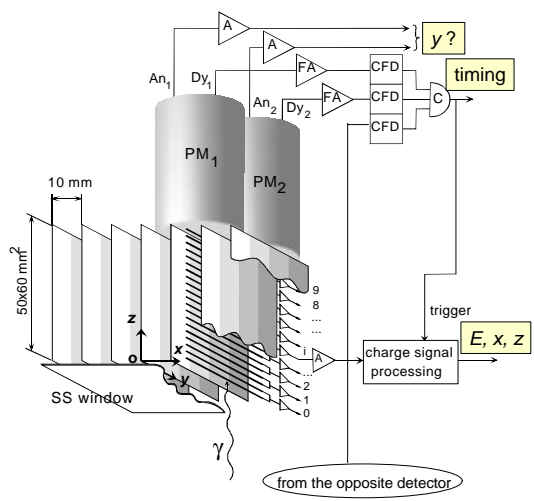

**Figure 2.** A schematic of the liquid xenon module for PET and its readout system.

The wires are 50 µm diameter spaced by 2.5 mm and connected in pairs, each pair being read individually with a low-noise charge sensitive preamplifier. This approach aims to keep the drift field reasonably uniform without doubling the number of readout channels. To reduce the input capacitance, the preamplifiers were installed directly on the airside of the chamber feedthroughs and operated at the liquid xenon temperature. We used the low temperature version of TOTEM 2.2 hybrid preamplifiers [13]. The input noise measured at –100ºC was about 350 electrons, r.m.s.

The scintillation is detected with two PMTs partially immersed into the liquid, viewing the cells from the top. The Hamamatsu R1668 1-1/8" photomultipliers with quartz window were used.

The readout and data acquisition systems were described in detail in [14].

*3.3  Use of scintillation*

As a result of the interaction of a γ-ray with liquid xenon a short scintillation pulse is produced. For a deposited energy of 511 keV, about 22,000 VUV photons are emitted. Liquid xenon scintillation is characterised with three time constants of 3 ns, 27 ns and 45 ns, the latter being associated with the recombination process and the other two components with direct excitation. In the presence of electric field, the recombination component is suppressed thus resulting in decrease of the light yield by a factor up to 3 and the component with decay time of 27 ns becomes dominant. Due to geometry of the cells, the efficiency of light collection to the PMTs was rather poor. No special measures were taken to improve the light reflection from the stainless steel cathode plates but even so, a reflectivity of ~20% has to be assumed to fit the dependence of the PMT signal amplitude on the distance between the photocathode and scintillation location. The number of photoelectrons per 511 keV varied from ~200 at the top of cell to ~30 at its bottom, i.e. by a factor of about 6 instead of ~100 that one expects from the solid angle.

The fast signal from photomultipliers is used to check for coincidence with the signal from a detector at the opposite side of the tomograph ring and also to trigger the data acquisition system. In our tests, we used a 2" $BaF_2$ crystal coupled to XP2020Q photomultiplier to detect the complementary γ-ray. The signals both from the $BaF_2$ and the chamber were fed to constant fraction discriminators with the threshold, for liquid xenon chamber, of about 1 photoelectron and, further, to time-to-amplitude converter and multichannel analyser. The coincidence time resolution of 1.3 ns (fwhm) was measured being, practically independent on the presence of electric field [15].

The efficiency of scintillation trigger has been studied in [16]. It was concluded that the minimum energy deposition reliably detected from the bottom of the cell is 50 keV, i.e. for the energy threshold of 50 keV the trigger efficiency is 100%.

### 3.4   Use of charge

In addition to scintillation, about 30,000 electron-ion pairs are created in liquid xenon by a 511 keV γ-ray, which deposits its total energy. From this amount, more than 90% of the electrons escape recombination at the electric field of 2 kV/cm and drift towards the collectors with the velocity of 2.3 mm/μs. Electron drift velocity is practically constant in liquid xenon for E>1 kV/cm. As the interaction instant is given by the scintillation, the electron collection time can be easily measured thus giving the interaction point position along *x*-direction (see Fig.2). The resolution measured with an α-source, which was placed on the cathode, was about 0.5 mm (*fwhm*) being mostly determined by non-uniformity of the electric field near the wires as shown in [17]. Contribution from the α-particle range is negligibly small (~50 μm for 5 MeV). Measurements with 511 keV γ-rays show that the resolution deteriorates to 0.8 mm due to contribution from the photoelectron range [18].

Fig.3 shows the signal pulse shape at the output of charge sensitive preamplifier. The lower trace in Fig.3,a corresponds to the channel (a pair of wires), in which the charge was collected, while the upper shows the signal read out the adjacent channel. The collecting channel is identified by reading the signal amplitude 3 μs after the trigger [14] thus allowing the measurement of *z*-position corresponding, in terms of PET, to the depth-of-interaction.

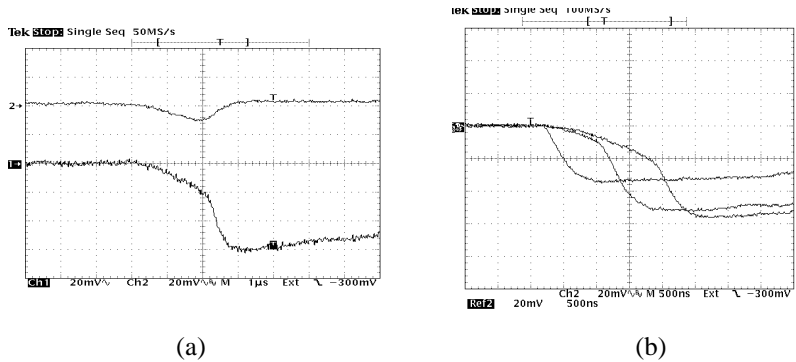

(a)                         (b)

**Figure 3**. (a) Charge signals induced in two adjacent channels. The lower trace corresponds to a pair of wires, to which the charge is collected; the upper trace shows the signal in the neighbour channel. (b) Charge signals induced in the collecting channel due to 511 keV γ-rays absorbed at different distances from the wire.

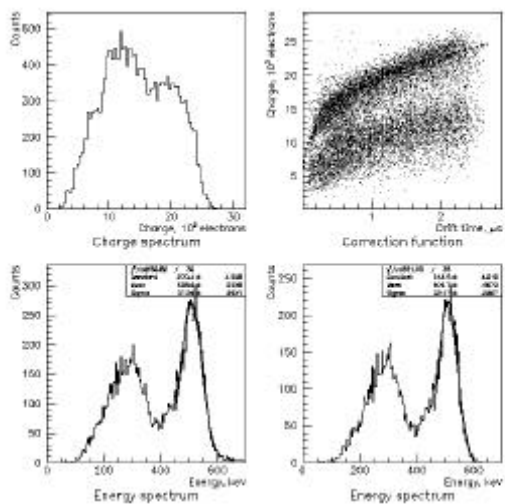

**Figure 4.** Amplitude spectrum of 511 keV γ-rays as measured (upper left); correlation plot of the collected charge as a function of drift time (upper right, a dash line shows the correlation function used for the amplitude correction); corrected spectra taking into account all events (bottom left) and only those with drift time >0.64 μs (bottom right).

Due to the low mobility of positive ions in the liquid only the electronic component of the signal can be observed with suitable integration time constants. This results in that the charge signal amplitude depends on the position where the ionisation has occurred (ballistic deficit) illustrated in Fig3,b. Since the standard solution for this problem (Frish grid) is difficult to implement due to limited space and inevitable increase of dead space in the detector (and, consequently, its efficiency), we used an alternative approach, which consists in the correction of the measured pulse amplitude using the information obtained from the rise time. From simulation, a correlation function between the pulse amplitude and ($x$, $y$)-position of the interaction point was found and the correction was applied to every detected signal [14]. The results are shown in Fig.4, in which one can see the amplitude spectrum due to 511 keV γ-rays as it is measured, the charge-time correlation plot and the corrected spectra. The obtained resolution is about 17% (*fwhm*) if all the events are taken into account and 15% in the case of rejecting the events with drift time shorter than 0.64 μs.

It was found that energy depositions as low as 50 keV can be detected with 100% efficiency along almost the whole cell except in two small regions at the cell top and bottom where the edge field effects result in losses of about 18% of the counts [16]. Improvement of the field uniformity at the edges can significantly reduce the losses.

### 3.5   Current improvements: the Mini-strip plate

An axial resolution of ~10 mm ($y$ in Fig.2) can be obtained by weighting the amplitudes of the PMT signals, which can be suitable, in some cases, for a PET tomograph with 2D slice by slice reconstruction, but it is not enough for using recently developed 3D reconstruction algorithms, which allow significant improvement of the system sensitivity.

In view of significant improvement of the axial resolution, a 2D charge readout was envisaged for replacing the wire plane and obtaining both the axial coordinate and depth-of-interaction ($y$ and $z$ in Fig.2). Hence, we developed what we call mini-strip plate, which besides providing good position resolution fulfils the specific requirements imposed by high purity environment and introduces a minimal dead space [19].

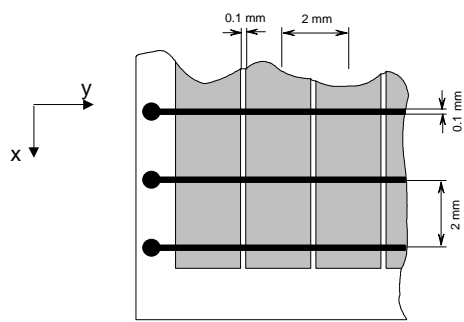

**Figure 4.** The mini-strip plate.

The mini-strip plate consists of two sets of metal strips deposited on both sides of a thin glass sheet, each strip being read individually. The strips on the side turned to the sensitive volume are thin (~0.1 mm) and widely spaced (2 mm) while those on the backside are, like a negative image, wide (1.9 mm) with a narrow gap (0.1 mm) between them. The two sets of strips are perpendicular to each other. A positive voltage is applied to the thin strips that operate as collectors and a slightly negative offset to the wide strips. Thus, the drifting electrons are collected to the thin strips and, at the same time, induce a charge signal on the back strips. The thinner the glass plate the larger the induced signal. For 0.5 mm D263 glass, for example, the amplitude of the induced charge signal is as high as 0.4 of that read out the collectors, for 0.25 mm it was computed to reach the value of about 0.5.

This system was tested with an α-source. The resolution ranging from 0.7 mm to 1.1 mm was obtained from the induction strips with the center of gravity method for the equivalent γ-ray energy of 220 keV and 120 keV, respectively ('equivalent energy' means the energy of a γ-ray that produces the same number of free electrons as that can be extracted from the α-particle track[1]). Extrapolating these results to $E_\gamma$=511 keV and adding the photoelectron range one gets the resolution of ~0.8 mm. The other coordinate was determined simply as the position of the strip, which collects the electrons, i.e. with the precision of 2 mm equal to the strip pitch. This is far enough for measuring depth-of-interaction in PET.

*3.6 Comparison with existing PET detectors*

Summarizing, the following parameters were obtained with the liquid xenon chamber for PET (all values are *fwhm*): time resolution 1.3 ns, axial and transaxial position resolutions of better than 1 mm, depth of interaction precision 5 mm (can be much better if necessary), energy resolution 17%, overall detection efficiency about 60% (being partly due to field non-uniformity that can be significantly improved; the probability for a γ-ray to interact with the liquid xenon can be increased, too, by increasing the detector thickness). These parameters are given in Tab.1 in comparison with those of two detection systems, one already existing and the other recently developed.

The major drawback of liquid xenon compared with heavy inorganic crystals is the smaller fraction of photoelectric absorption. In BGO, for example, the photo-fraction for 511 keV γ-rays is about 0.44 while it is only 0.22 for xenon. This means that for about 0.2 of all coincidences in BGO both gamma-photons are detected in the total absorption peak. For xenon, this fraction is 0.05. However, there is an important difference between a scintillator and the liquid xenon drift chamber in what concerns the capability of processing the events that suffer Compton scattering in the detector. In the scintillation detector these events are usually rejected as the two (or more) interaction points can not be distinguished and, therefore, such events result in a biased measured position. On the contrary, the liquid xenon drift chamber, which has effectively a voxel (3D-pixel) structure, allows these points to be identified and their position and energy deposition in each of them measured. Therefore, the reconstruction of the event topology is possible so that the first interaction in the detector can be found and its position used in the image reconstruction. It was shown that a significant fraction of the Compton events in the detector can be recovered and used for the image reconstruction [22]. The total deposited energy can also be measured thus allowing the rejection of scatter in the object (human body).

---

[1] Recombination along the track of α-particles in liquid xenon is such that only a few percent of the electron-ion pairs escape recombination at E<10 kV/cm.

Table 1. Comparison of the liquid xenon detector with scintillation crystal systems

|  | PETYA | BGO block detector [20] | LSO block detector (CTI) [21] |
|---|---|---|---|
| Time resolution | 1.3 ns | 2 ns | 1.5 ns |
| Position resolution | 0.8×0.8 mm$^2$(*) | 5×5 mm$^2$ | 2×2 mm$^2$ |
| Interaction depth resolution | 2 to 5 mm | None | 7.5 mm |
| Energy resolution | 15% to 17% | 20% | 14% to 20% (**) |
| Efficiency | 60% | 80% | not quoted |
| Dead time | 50 µs·cm$^2$ | 25 µs·cm$^2$ | not quoted |

∗ $\Delta x \times \Delta y$; $\Delta x$ - from the drift time measurement; $\Delta y$ – obtained with the center of gravity method with the mini-strip plate (extrapolated from the measurements with α-source and convoluted with the photoelectron range)
** for a single crystal

## 4   Liquid xenon gamma camera

In the experiments with α-particles it was shown that the mini-strips plate allows reliable detection of the events with the number of electrons extracted from the track as low as 7,000 that corresponds to the γ-ray energy of about 120 keV. Thus, a gamma camera with ionisation readout for medical diagnostic, which typically uses γ-rays of 140 keV, seemed to be feasible. A small prototype of such device, schematically shown in Fig.5, has been produced and is currently under tests. It uses the mini-strip plate with an active area of 50x50 mm$^2$, made of D263-glass 0.55 mm thick, with the strip sets identical to those descried in the previous section. 22 strips are read out each side of the plate with two integrated multichannel circuits that include charge sensitive preamplifiers, discriminators and sample and hold circuits [23]. The input noise of ~100 electrons, r.m.s., was measured at the liquid xenon temperature and zero capacitance. With the inputs connected to the strips, the noise increases to ~240 and ~340 electrons for the collecting and induction strips, respectively.

The first tests with 122 keV γ-rays has shown that a resolution better than 2 mm (*fwhm*) can be obtained with this method. We refer to [24] for further details.

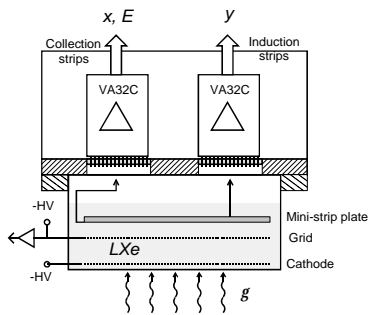

**Figure 5.** Schematic view of the mini-strip 2D camera.

## 5  Acknowledgements

This work was done in the framework of the project POCTI /SAU/1342/95; one of the authors was supported by the fellowship PRAXIS XXI /BD/3892/96, both from Fundação para a Ciência e Tecnologia, Portugal.